\documentclass{article}
\usepackage{epsf}
\usepackage{amsfonts}
\usepackage{times}
\newcommand{\braket}[2]{\langle #1 | #2 \rangle}
\newcommand{\ket}[1]{\left | \, #1 \right \rangle}

\newcommand{\bra}[1]{\left \langle #1 \, \right |}

\newcommand{\vpols}{\mbox{$\updownarrow$}}
\newcommand{\hpols}{\mbox{$\leftrightarrow$}}

\newcommand{\ppols}{\mbox{\small
\,\mbox{$\nearrow$}\llap{\mbox{$\swarrow$}}\,}}

\newcommand{\qpols}{\mbox{\small
\,\mbox{$\searrow$}\llap{\mbox{$\nwarrow$}}\,}}

\newcommand{\rpols}{\mbox{\raisebox{-0.66ex}{$\triangleleft$}$\!\!\supset$}}
\newcommand{\lpols}{\mbox{$\subset\!\!$\raisebox{-0.66ex}{$\triangleright$}}}

\newcommand{\Tr}{{\mathrm{Tr}}}

\begin{document}
\vspace{1in}

\begin{center}
{\Large Capacities of Quantum Channels and How to Find Them
}\\
{\large Peter W. Shor}\\
{AT\&T Labs -- Research} \\
Florham Park, NJ 07922
\end{center}

\noindent
{\bf Abstract:} We survey what is known about the information
transmitting capacities of quantum channels, and give a proposal 
for how to calculate some of these capacities using linear
programming.

\section{Introduction}
In this paper, we discuss the capacity of quantum channels.  Information
theory says that the capacity of a classical channel is
essentially unique, and is representable as a single numerical quantity,
which gives the amount of information that can be transmitted asymptotically 
per channel use \cite{Shannon48,Cover}. Quantum channels, unlike
classical channels, do not have a single numerical quantity which can
be defined as their capacity for transmitting information.  Rather, 
quantum channels appear to have at least four different natural 
definitions of capacity, depending on the auxiliary resources allowed, 
the class of protocols allowed, 
and whether the information to be transmitted is classical or quantum.

In this paper, we first introduce the background necessary for
understanding the capacity of quantum channels, and   
then define several
capacities of these channels.
For two of these channel capacities,  
we sketch possible techniques for
computing them which we believe will be more efficient than techniques
currently used.
These capacities are both reducible to
optimization problems over matrices.  We believe that a combination of
linear programming techniques, including column generation, and non-linear
optimization will provide a more efficient method for calculating these
capacities.  Unfortunately, at the time of 
writing this paper, I have not yet tested these techniques 
experimentally.  Since I cannot prove that these techniques are 
efficient,
the proof of this pudding must be in the computing, and is thus not yet
demonstrated.  We hope to test these techniques in the near future.

To date, the means used
for numerical computations of quantum channel capacities have
been fairly straightforward, often using gradient descent
techniques \cite{OsNa}.  More research has been done 
on the calculation of 
the entanglement of formation \cite{Tucci, Aud2}, a related problem
\cite{MSW}.  None of these programs have used combinatorial optimization
techniques.  For one of the capacities discussed in 
this paper---the entanglement-assisted capacity---this technique may 
be fairly efficient, as this capacity has a single local
optimum which is also a global optimum.  For two other
capacities discussed in this paper---the $C_{1,1}$ and $C_{1,\infty}$ 
capacities---I propose techniques involving linear 
programming that could be used for the capacity computation, 
and which I suspect
are much more efficient than straightforward optimization.  For another
capacity---the one-way quantum capacity---there are multiple local 
maxima in the optimization problem, and we need to determine the global 
maximum.  In this case, unfortunately,
although hill climbing does not seem like it would be efficient,
I do not have any alternative techniques to suggest.

This paper originates in my research investigating the 
capacities of a quantum channel \cite{ping-pong}.  In order to show that a 
certain
channel capacity (which I do not deal with in this paper; it is less
natural than the capacities covered here) lies strictly between 
two other channel capacities, I
needed to calculate some of these capacities.  Specifically, I needed 
to calculate
what I call the $C_{1,1}$ capacity of a fairly simple quantum channel.
I realized that this was a problem which could be solved numerically
using linear programming, and I used this technique to obtain a
picture of the $C_{1,1}$ capacity landscape which was satisfactory for 
my application.  During this computation,
it became clear that a better way to solve this problem would be to use
column generation techniques to make the linear program more efficient, 
and that these would furthermore also be
useful for computing other capacities of quantum channels.   I have 
not yet had time to experimentally test these new techniques (rather, 
my program started with enough columns to ensure obtaining
a close approximation of the capacity; this would be an enormous waste 
of resources for larger problems, but for my purposes
it was quite adequate).  This paper will explain
the column generation technique.  I will try to make it comprehensible both
to researchers with background in mathematical programming and to researchers
with background in quantum information theory.  Those wishing more
background information on linear programming, on quantum computing and 
information, or on classical information theory can find them in
textbooks such as  
\cite{Chvatal,Nielsen-Chuang,Cover}
More specifically, I will
give proposals for how to compute two capacities for carrying classical
information over a quantum channel: namely, the $C_{1,1}$ capacity and
the $C_{1,\infty}$ capacity.  These techniques should also work for computing
a formula that I conjecture gives the classical entanglement-assisted 
capacity with limited entanglement;
this extrapolates between the $C_{1,\infty}$ capacity and the 
entanglement-assisted capacity.   
The description of quantum information
theory and capacities contained in here is largely taken from the paper
\cite{gafa}.

\section{Quantum Information Theory}

The discipline of information theory was founded by Claude Shannon
in a truly remarkable paper \cite{Shannon48} which laid down the foundations
of the subject.  
We begin with a quote from this paper
which a nutshell summarizes one of the main concerns of 
information theory:
\begin{quote}
{\em
The fundamental problem of communication is that of
reproducing at one point either exactly or approximately a
message selected at another point.
}
\end{quote}
This paper proposed the definition of the capacity of a classical channel
as the amount of information per channel use that can be transmitted
asymptotically in the limit of many channel uses,
with near perfect reproduction at the receiver's end, and gave a simple
and elegant formula for the capacity. 
Here, the information is the logarithm (base 2) of the number of messages, 
in other words the number of classical bits that can be transmitted by
the channel.

The definition of quantum channel capacity is motivated largely by the same 
problem, with the difference being that either the method of reproduction or 
the message itself involves fundamentally quantum effects.  
For many years, information theorists either ignored quantum effects
or approximated them so as to make them susceptible to classical analysis; it
was only in the last decade or so that the systematic study of
quantum information theory began.

Shannon's original paper set forth two coding theorems which form the
foundation of the field of information theory.  The first is the source
coding theorem, which gives a formula for how much a random information
source can be compressed.  The second is the channel coding theorem,
which gives a formula for how much redundancy must be added to a message
in order to accurately reproduce it after sending the information through
a noisy channel.  

\section{Shannon theory}

Shannon's 1948 paper \cite{Shannon48} contained two theorems for 
which we give quantum analogs.  The first of these is the 
{\em source coding} theorem, which gives a formula for how much a source
emitting random signals can be compressed, while still permitting the
original signals to be recovered with high probability.  
Shannon's source coding theorem states that $n$ outputs of a source~$X$ 
can be compressed to length $n H(X) + o(n)$ bits, and restored to 
the original with high probability, where $H$ is the entropy function.
For a probability distribution with probabilities $p_1$, $p_2$, $\ldots$,
$p_n$, the entropy $H$ is
\begin{equation}
H(\{p_i\}) = \sum_{i=1}^n -p_i \log p_i,
\end{equation}
where information theorists generally take the logarithm base 2
(thus obtaining bits as the unit of information).  

The second of these 
theorems is the {\em channel coding} theorem, which states that with
high probability, $n$ uses of
a noisy channel $N$ can communicate $Cn - o(n)$ bits reliably, where 
$C$ is the channel capacity given by 
\begin{equation}
C = \max_{p(X)} \ I(X; N(X))
\label{classical-cap}
\end{equation}
Here the maximum is taken over all probability distributions on inputs
$X$ to the channel, and $N(X)$ is the output of the channel given input
$X$.  The {\em mutual information} $I$ between two random variables $X$ and
$Y$ is defined as:
\begin{eqnarray}
I(X;Y) &=& H(Y) - H(Y|X) \label{output-given-input}\\
&=& H(X) + H(Y) - H(X,Y), \label{input-plus-output-less-joint}
\end{eqnarray}
where $H(X,Y)$ is the entropy of the joint distribution of $X$ and $Y$, and 
$H(Y|X)$ is the conditional
entropy of $Y$, given $X$. That is, if the possible values
of $X$ are $\{X_i\}$, then the conditional entropy is
\begin{equation}
H(Y|X) = \sum_i \Pr(X=X_i) \, H(Y|X=X_i).
\end{equation}
There is an efficient algorithm, the {\em Arimoto-Blahut algorithm,}
for calculating the capacity (\ref{classical-cap}) of a classical channel
\cite{Arimoto,Blahut,Cover,Sayir}.

When the formula for mutual information is extended to the quantum case,
two generalizations have been found that both give capacities
of a quantum channel, although these capacities differ in both the 
resources that the sender and receiver have available and the operations they
are permitted to carry out.  One of these formulae generalizes
the expression (\ref{output-given-input}) and the other the expression
(\ref{input-plus-output-less-joint}); these expressions are equal
in the classical case.

For classical channels, there are a number of extra resources which one 
might imagine could increase their capacity.  These include
a feedback channel from the receiver to the sender and shared
randomness between the sender and the receiver.  It turns out that
neither of these resources actually does increase the capacity of a 
classical channel.  For quantum channels, however, the situation is 
different.  In this case, one of the resources that we must consider
is entanglement.  An entangled pair of quantum states consists of two states
which are non-classically correlated.  To parties who share such a
pair of states cannot use them to transmit information, but can use
them to obtain a shared random variable.
It turns out that shared entanglement between
the sender and receiver can be used to increase the transmission capacity 
of a quantum channel.
When the capacity of a quantum channel for transmitting quantum 
information is considered, things become even more complicated.  In 
this case, a classical side channel can increase the capacity of a
quantum channel to transmit quantum information, even though no quantum
information can be transmitted by a classical channel.  

\section{Quantum mechanics}
Before we can start talking about quantum information theory, I need to give
a brief description of some of the fundamental principles of quantum
mechanics.  The first of these principles
that we present is the {\em superposition principle}.
In its most basic form, this principle says that if a quantum
system can be in one of two distinguishable states $\ket{x}$ and
$\ket{y}$, it can be in any state of the form 
$\alpha \ket{x} + \beta \ket{y}$, where $\alpha$ and $\beta$ are 
complex numbers with $|\alpha|^2 + |\beta|^2=1$.  Here $\ket{\cdot}$ 
is the {\em bra-ket} notation that physicists use for a quantum state; 
we will occasionally be using it in the rest of this paper.  Recall that
we assumed
that $\ket{x}$ and $\ket{y}$ were distinguishable, so there must 
conceptually be
some physical experiment which distinguishes them (this 
experiment need not be performable in practice).  The principle says
further that if we perform this experiment, we will observe the state $\ket{x}$
with probability $|\alpha|^2$ and $\ket{y}$ with probability 
$|\beta|^2$.  Furthermore, after this experiment is performed, if
state $\ket{x}$ (or $\ket{y}$) is observed the system will thereafter behave
in the same way as
it would have had it originally been in state $\ket{x}$
(or $\ket{y}$).

Mathematically, the superposition principle says that the states of a
quantum system are the unit vectors of a complex vector space, that
two orthogonal vectors are distinguishable, and that measurement projects
the state onto one of an complete orthonormal set of basis vectors.  
In accordance with physics usage, we will represent quantum states
by column vectors.  
The Dirac {\em bra-ket} 
notation denotes a column vector by $\ket{v}$ (a {\em ket}) and its
Hermitian transpose (i.e., complex conjugate transpose) by
$\bra{v}$ (a {\em bra}).  
The inner product between two vectors, $v$ and $w$, is denoted 
$\braket{w}{v} = w^\dag v$, here we define $X^\dag$ (whether $X$ is a
vector or matrix) to be the Hermitian transpose of $X$.
Multiplying a quantum state vector by a complex phase
factor (a unit complex number) does not change any properties
of the system, so mathematically the state of a quantum system is 
a point in projective complex space. 
Unless otherwise stated, however, we will represent quantum states
as unit vectors in some complex vector space ${\mathbb C}^d$.

We will be dealing solely with finite dimensional vector spaces.  
For an introductory paper, quantum information theory
is already complicated enough in finite dimensions 
without introducing
the additional complexity of infinite-dimensional vector spaces.
Many of the theorems we discuss do indeed generalize
naturally to infinite-dimensional spaces.

A {\em qubit} is a two-dimensional quantum system.  Probably the most widely
known qubit is the polarization of a photon, and we will thus
be using this 
example.  For the polarization of a photon,
there can only be two distinguishable states.  If one sends a photon 
through a birefringent crystal, it will take one of two paths, depending 
on its polarization.  By re-orienting this crystal, these two distinguishable
polarization states can be chosen to be horizontal and vertical, or 
right diagonal and left diagonal.  In accordance
with the superposition principle, each of
these states can be expressed as a complex combination of basis states
in the other basis.  For example,
\begin{eqnarray*}
 \ket{\ppols} &=& \frac{1}{\sqrt{2}} \ket{ \hpols  } +
\frac{1}{\sqrt{2}} \ket{ \, \vpols \, }\\
 \ket{\qpols} &=& \frac{1}{\sqrt{2}} \ket{ \hpols } -
\frac{1}{\sqrt{2}} \ket{ \, \vpols \, }\\
 \ket{\,\rpols} &=& \frac{1}{\sqrt{2}} \ket{ \hpols } +
\frac{i}{\sqrt{2}} \ket{ \, \vpols \, }\\
 \ket{\,\lpols} &=& \frac{1}{\sqrt{2}} \ket{ \hpols } -
\frac{i}{\sqrt{2}} \ket{ \, \vpols \, }
\end{eqnarray*}
Here, $\ket{\,\rpols}$ and $\ket{\,\lpols}$ stand for right and left
circularly polarized light, respectively;  these are another pair of
basis states for the polarization of photons.  For a specific
example, when diagonally
polarized photons are put through a birefringent crystal oriented in the
$\vpols, \hpols$ direction, half of them will behave like vertically
polarized photons, and half like horizontally polarized photons; thereafter,
these photons will indeed have these polarizations.

If you have two quantum systems, their joint state space is the
tensor product of their individual state spaces.  For example, the 
state space of two qubits is~${\mathbb C}^4$ and of three qubits
is ${\mathbb C}^8$.  The high dimensionality of the space for $n$ qubits, 
${\mathbb C}^{2^n}$,
is one of the places where quantum computation attains its power.

The polarization state space of two photons has as a basis the four states
\[
\ket{\vpols\,\vpols\,}, \quad \ket{\vpols\,\hpols\,}, 
\quad \ket{\hpols\,\vpols\,}, \quad \ket{\hpols\,\hpols\,}.
\]
This state space includes states such as an EPR (Einstein, Podolsky,
Rosen) pair of photons
\begin{equation}
\frac{1}{\sqrt{2}}(\ket{\vpols\,\hpols\,} - \ket{\hpols\,\vpols\,}) =
\frac{1}{\sqrt{2}}(\ket{\ppols\,\qpols\,} - \ket{\qpols\,\ppols\,}),
\label{EPR}
\end{equation}
where neither qubit alone has a definite state, but which 
has a definite state when considered as a joint system
of two qubits.  In this state,
the two photons have orthogonal polarizations in whichever basis they
are measured in.  Bell \cite{Bell-theorem} showed that
the outcomes of measurements on the photons
of this state cannot be reproduced by
joint probability distributions which give probabilities for
the outcomes of 
all possible measurements, and in which each 
of the single photons has a definite probability distribution for the
outcome of measurements on it, independent of the measurements which are
made on the other photon \cite{Bell-theorem,GHSZ}.  
In other words, there cannot be any set of hidden variables associated
with each photon that determines the probability distribution obtained
when this photon is measured in any particular basis.
Two quantum systems such as an EPR pair which are non-classically
correlated are said to be {\em entangled} \cite{Bruss}.

The next fundamental principle of quantum mechanics we discuss is the
{\em linearity principle.}  This principle states that an isolated quantum 
system undergoes linear evolution.  Because the quantum systems we are 
considering are finite dimensional vector spaces, a linear evolution of 
these can be described by multiplication by a matrix.  It is fairly easy
to check that in order to make the probabilities sum to one, we must
restrict these matrices to be unitary (a matrix $U$ is unitary if 
$U^\dag = U^{-1}$; unitary matrices are those complex matrices which
take unit vectors to unit vectors).  

Although many elementary treatments of quantum mechanics restrict themselves 
to pure states (unit vectors), for quantum information theory we need to
treat probability distributions over quantum states.  These naturally
give rise to objects called density matrices.  For an $n$-dimensional
quantum state space, a {\em density matrix}
is an $n \times n$ Hermitian trace-one
positive semidefinite matrix.  

Density matrices arise naturally from quantum states in two ways.
The first way in which density matrices arise is from probability distributions
over quantum states.  
A rank one density matrix $\rho$ corresponds to the pure state $v$
where $\rho = v v^\dag$.  (Recall $v^\dag$ was the 
Hermitian transpose of $v$.)  
Suppose that we have a system which is in state
$v_i$ with probability $p_i$.  The corresponding density matrix is
\begin{equation}
\rho = \sum_i p_i v_i v_i^\dag.
\end{equation}

An important fact about density matrices is that the density matrix
for a system gives as much 
information as it is possible to obtain
about experiments performed on the system.  
That is, any two systems with the same density matrix $\rho$ 
cannot be distinguished by experiments, provided that no extra side 
information is given about these systems.

The other way in which density matrices arise is through disregarding part 
of an entangled quantum state.  Recall that two systems in an entangled pure 
state have a definite quantum state when considered jointly, but that neither
of the two systems individually can be said to have a definite state.  
The state of either of these systems considered separately
is naturally represented by a density matrix.
Suppose that 
we have a state $\rho_{AB}$ on a tensor product system ${\cal H}_A
\otimes {\cal H}_B$.  If we can only see the first part of the system,
this part behaves as though it is in the state $\rho_A = \Tr_B \rho_{AB}$.
Here, $\Tr_B$ is the partial trace operator.  Consider a joint system
in the state
\begin{equation}
\rho_{AB} = \left(
\begin{array}{ccc}
B_{11}&  B_{12}& B_{13}\\
B_{21}&  B_{22}& B_{23}\\
B_{31}&  B_{32}& B_{33}
\end{array}
\right).
\end{equation}
In this example, the dimension of ${\cal H}_A$ is 3 and the dimension of 
${\cal H}_B$ is the size of the matrices $B_{ij}$.
The partial trace of $\rho_{AB}$, tracing over ${\cal H}_A$, is 
\begin{equation}
\mathrm{Tr}_A \ \rho_{AB} =
B_{11} +  B_{22} + B_{33}
\end{equation}
Although the above formula also
determines the partial trace when we trace over ${\cal H}_B$, through a
permutation of the coordinates, it is instructive to give this explicitly:
\begin{equation}
\mathrm{Tr}_B \ \rho_{AB} = \left(
\begin{array}{ccc}
\mathrm{ Tr} \,B_{11}&  \mathrm{ Tr} \,B_{12}& \mathrm{ Tr} \,B_{13}\\
\mathrm{ Tr} \,B_{21}&  \mathrm{ Tr} \,B_{22}& \mathrm{ Tr} \,B_{23}\\
\mathrm{ Tr} \,B_{31}&  \mathrm{ Tr} \,B_{32}& \mathrm{ Tr} \,B_{33}
\end{array}
\right).
\end{equation}

The final ingredient we need before we can
start explaining quantum information theory is a
{\em von Neumann measurement}.  We have seen examples of this process before, 
while explaining the superposition principle; 
however, we have not yet given the general mathematical
formulation of a von Neumann measurement.  Suppose that we have an 
$n$-dimensional
quantum system~${\cal H}$.  A von Neumann measurement corresponds to
a complete set of orthogonal subspaces $S_1$, $S_2$, $\ldots$, $S_k$ 
of ${\cal H}$. 
Here, complete means that the subspaces $S_i$ span the space ${\cal H}$, so
that $\sum_i \dim S_i = n$.  Let $\Pi_i$ be the projection matrix
onto the subspace $S_i$.  If we start with a density matrix $\rho$, 
the von Neumann measurement corresponding to the set of subspaces~$\{S_i\}$ 
projects
$\rho$ into one of the subspaces $S_i$.  Specifically, it projects
$\rho$ onto the $i$'th subspace with probability $\Tr \ \Pi_i\rho$, 
the state after the projection being 
\[
\frac{1}{\Tr\, \Pi_i \rho } \Pi_i \rho \Pi_i,
\]
where we have renormalized the projection to have trace 1. A special
case that is often encountered is when the $S_i$ are all one-dimensional, 
so that $S_i = w_i w_i^\dag$, and the vectors $w_i$ form an orthogonal
basis of ${\cal H}$.  Then, a vector $v$ is taken to $w_i$ with
probability
$|w_i^\dag v|^2$, and a density matrix $\rho$ is taken to $w_iw_i^\dag$ with
probability $w_i^\dag \rho w_i$.

\section{Von Neumann entropy}
We are now ready to consider quantum information theory.  We will start by
defining the entropy of a quantum system.  To give some intuition for
this definition, we first consider some special cases.  Consider $n$
photons, each being
in the state $\ket{\vpols}$ or $\ket{\hpols}$ with probability
$\frac{1}{2}$.  Any two of these states are completely distinguishable.
There are thus $2^n$ equally probable states of the system, and the entropy
is $n$ bits.  This is essentially a classical system.

Consider now $n$ photons, each being
in the state $\ket{\vpols}$ or
$\ket{\ppols}$ with probability~$\frac{1}{2}$.  These states are not 
completely distinguishable, so there are effectively considerably
less than $2^n$ states, and the entropy should intuitively
be less than $n$ bits.

By thermodynamic arguments involving the increase in entropy associated
with the work extracted from a system, von Neumann deduced that the 
{\em (von Neumann) entropy} 
of a quantum system with density matrix $\rho$ should be 
\begin{equation}
H_\mathrm{vN} (\rho) = - \Tr \rho \log \rho.
\end{equation}
Recall that $\rho$ is positive semidefinite, so that $-\Tr \rho \log \rho$
is well defined.  If $\rho$ is expressed in coordinates in which it
is diagonal with eigenvalues $\lambda_i$, then in these coordinates 
$- \rho \log \rho$ is diagonal with
eigenvalues $-\lambda_i \log \lambda_i$.  We
thus see that 
\begin{equation}
H_\mathrm{vN}(\rho) =  H_\mathrm{Shan}(\lambda_i),
\end{equation}
so that the von Neumann entropy of a density matrix is the Shannon
entropy of the eigenvalues.  (Recall $\Tr \rho = 1$, so that
$\sum_i \lambda_i = 1$.)  This definition is easily seen to agree
with the Shannon entropy in the classical case, where all the states 
are distinguishable.

\section{Source coding}

Von Neumann developed the above definition of entropy for thermodynamics.
One can ask whether this is also the correct definition of entropy
for information theory.  We will first give the example of quantum
source coding \cite{JS94, Schumacher}, 
also called {\em Schumacher compression,} for which 
we will see that it is indeed the right definition.  We consider
a memoryless
quantum source that at each time step
emits the pure state $v_i$ with probability $p_i$.
We would like to encode this signal in as few qubits as possible, and send
them to a receiver who will then be able to reconstruct the original state.
Naturally, we will not be able to transmit the original state flawlessly.
In fact, the receiver cannot even reconstruct the
original state absolutely
perfectly most of the time (this is the corresponding requirement
in classical information theory).  
Unlike classical signals, quantum states are not 
completely distinguishable theoretically,
so reconstructing the original state most of the time is too
stringent a requirement.
What we require is that the receiver be able to reconstruct a state
which is almost completely indistinguishable from the original state
nearly all the time.  For this we need a measure of indistinguishability;
we will use a measure called {\em fidelity}.  Suppose that the
original signal is a vector 
\[
u = v_1 \otimes v_2 \otimes \ldots \otimes v_n.
\]
Then the fidelity between the signal $u$ and the output $\rho$ (which is
in general a mixed state, i.e., a density matrix, on $n$ qubits) is 
$F = u^\dag \rho u$.
The average fidelity is this fidelity $F$ averaged over $u$.  If the
output is also a pure state $v$, the fidelity 
$F = u^\dag v v^\dag u = |u^\dag v|^2$.  If the input is a pure state,
the fidelity
measures the probability of success of a test which
determines whether the output
is the same as the input.  
If both the output state $\rho_{\mathrm{out}}$ and the input state
$\rho_{\mathrm{in}}$ are mixed states,
the fidelity is defined 
\[
\Tr \sqrt{\rho_{\mathrm{out}}^{1/2}\rho_{\mathrm{in}}\rho_{\mathrm{out}}^{1/2}},
\]
an expression which, despite its appearance, is symmetric in 
$\rho_{\mathrm{in}}$ and $\rho_{\mathrm{out}}$ \cite{fidelity-def}. In
the case where either $\rho_{\mathrm{out}}$ or $\rho_{\mathrm{in}}$ is pure,
this is equivalent to the previous definition, and for mixed states it is
a relatively simple expression which gives an upper bound on the 
probability of distinguishing these two states.  

Before I can continue to sketch the proof of the quantum source coding
theorem, I need to review the proof of the classical source coding theorem.
Suppose we have a memoryless source, i.e., a source $X$ that at each
time step emits the $i$'th signal type $S_i$ with probability $p_i$,
and where the probability distribution for each signal is independent 
of the previously emitted signals.  The idea behind classical
source coding is to show that with high probability, the source emits
a {\em typical sequence.}  Here a sequence of length $n$ is 
defined to be {\em typical} 
if it contains approximately $n p_i$ copies of the signal $S_i$ for
every~$i$.\footnote{Strictly speaking, this is the definition of
frequency-typical sequences.  Their is a related but distinct 
definition of entropy-typical sequences and subspaces, which can 
also be used in many of these proofs.}  The
number of typical sequences is only $2^{n H(X) + o(n)}$.  These can thus
be coded in $n H(X) + o(n)$ bits.

The tool that we use to perform Schumacher compression is that of
{\em typical subspaces}.  Suppose that we have a density matrix 
$\rho \in {\cal H}$, where ${\cal H} = {\mathbb C}^k$,
and we take the tensor product of $n$ copies of $\rho$ in the space 
${\cal H}^n$, i.e., we take $\rho ^{\otimes n} \in {\mathbb C}^{nk}$.
There is a typical subspace associated with $\rho^{\otimes n}$.  Let 
$\hat{v}_1$, $\hat{v}_2$, $\ldots$, $\hat{v}_k$ be the eigenvectors
of $\rho$ with associated eigenvalues 
$\lambda_1$, $\lambda_2$, $\ldots$, $\lambda_k$.  Since $\Tr \rho =1$,
these $\lambda_i$ form a probability distribution.  Consider typical 
sequences of the eigenvectors $\hat{v}_i$, where $\lambda_i$ is
the probability of choosing $\hat{v}_i$.  
A typical sequence can be turned into a
quantum state in ${\cal H}^{\otimes n}$ by taking
the tensor products of its elements.  That is,
if a typical sequence is 
$\hat{v}_{i_1}$, $\hat{v}_{i_2}$, $\ldots$, $\hat{v}_{i_n}$, the 
corresponding quantum state is
$w = \hat{v}_{i_1}\otimes \hat{v}_{i_2}\otimes \ldots\otimes \hat{v}_{i_n}$.
The typical subspace ${\cal T}$
is the subspace spanned by typical sequences of the
eigenvectors.  The subspace $\cal T$ has dimension equal to the number
of typical sequences, or $2^{H_\mathrm{vN} (\rho) n + o(n)}$.

We can now explain how to do Schumacher compression.  Suppose we wish
to compress a source emitting $v_i$ with probability $p_i$.  Let the
typical subspace corresponding to $\rho^{\otimes n}$ be $\cal T$, 
where $\rho = \sum_i p_i v_i v_i^\dag$ is the density matrix for the
source, and where we are using a block length $n$ for our compression
scheme.  We take the vector
$u= v_{i_1} \otimes v_{i_2} \otimes \ldots \otimes v_{i_n}$ and make
the von Neumann measurement that projects it into either ${\cal T}$ 
or ${\cal T}^\perp$.  If $u$ is projected onto ${\cal T}$, we send
the results of this projection to the receiver; this can be done with 
$\log \dim {\cal T} = n H_\mathrm{vN}(\rho) + o(n)$ qubits.  If $u$
is projected onto ${\cal T}^\perp$, our compression algorithm has
failed and we can send anything; this does not degrade the fidelity 
of our transmission much, because this is a low probability event.

Why did this work?  We give a brief sketch of the proof.
The main element of the proof is to show that the 
probability that we project $u$ onto ${\cal T}$ approaches 1 as $n$ goes 
to $\infty$.  
This probability is $u^\dag \Pi_{\cal T} u$.  If
this probability were exactly 1, then
${u}$ would necessarily be in ${\cal T}$, and we would have
noiseless compression.  If this probability is close to 1, then
$u$ is close to the subspace ${\cal T}$, and so $u$ has high
fidelity with the projected vector $\Pi_{\cal T} u$.  Suppose
the probability that the state ${u}$ is projected onto
${\cal T}$ is $1-\epsilon$.  Then $u^\dag \, \Pi_{\cal T}\,  u = 1-\epsilon$
and the fidelity between the original 
state ${u}$ and the final state is
$|\braket{u}{\Pi_{\cal T}\,u}|^2 = (1-\epsilon)^2$.

Now, recall that if two density matrices are equal, the outcomes of any 
experiments performed on them have the same probabilities.  Thus,
the probability that the source $v_i$ with probabilities $p_i$ is projected
onto the typical subspace is the same as for the source $\hat{v}_i$ with
probabilities $\lambda_i$, where $\hat{v}_i$ and $\lambda_i$ are the
eigenvalues and eigenvectors of $\rho = \sum_i p_i v_i v_i^\dag$.  Because
the $\hat{v}_i$ are distinguishable, this is essentially the classical 
case, and $w$ is in the typical subspace 
exactly when the sequence of $\hat{v}_i$ is a typical sequence.  
We then
know from the classical theory of typical sequences that 
$w = \hat{v}_{i_1}\otimes \hat{v}_{i_2}\otimes \ldots \otimes \hat{v}_{i_k}$
is in the typical subspace at least $1-\epsilon$ of the time, completing
the proof.

\section{Accessible information and the $C_{1,1}$ capacity}
\label{section-examples}

The next concept we consider
is that of {\em accessible information}.  Here, we again 
have a source emitting state $\sigma_i$ with probability $p_i$.  Note that now,
the states $\sigma_i$ emitted may be density matrices rather than pure states.
We will ask a different question this time.  We now want to obtain as much
information as possible about the sequence of signals emitted by the source.
This is called the {\em accessible information} of the source.  
That is, the accessible information is the maximum over all measurements
of the mutual information 
$I(X;Y)$ where $X$ is the random variable telling which signal $\sigma_i$ 
was emitted by the source, and $Y$ is the random variable giving the 
outcome of a measurement on $\sigma_i$.  This gives the capacity of a 
channel where at each time step the sender must choose one of the states 
$\sigma_i$ to send, and must furthermore choose $\sigma_i$ a fraction
$p_i$ of the time; and where the receiver must choose a fixed measurement 
that he will make on every signal received.

To find the accessible information, we 
need to maximize over all measurements.  For this, we need to
be able to characterize all possible quantum measurements.  It turns
out that von Neumann
measurements are not the most general class of quantum measurements; the most
general measurement is called a
{\em positive operator valued measure,} or {\em POVM}.   
One way to describe these is as von
Neumann measurements on a quantum space larger than the original space;
that is, by supplementing the quantum state space by an {\em ancilla}
space and taking a von Neumann measurement on the joint state space.

We now give a more effective, but equivalent, characterization of POVM's.
For simplicity,
we restrict our discussion to POVM's with a finite number of distinct
outcomes; these turn out to be sufficient for studying capacities of finite 
dimensional channels.
A POVM can be defined by a set of positive semidefinite matrices $E_i$
satisfying $\sum_i E_i = I$.  If a quantum system has
matrix $\rho$, then the probability of the $i$'th outcome
is 
\begin{equation}
p_i = \Tr ( E_i \rho)
\end{equation}
For a von Neumann measurement, we take $E_i = \Pi_{S_i}$, the projection matrix
onto the $i$'th orthogonal subspace $S_i$.  The condition 
$\sum_i \Pi_{S_i} = I$ is equivalent to the requirement 
that the $S_i$ are orthogonal and span the whole state space.
To obtain the maximum information from a POVM, we can assume that the 
$E_i$'s are pure states; if there is
an $E_i$ that is not rank one, then we
can always achieve at least as much accessible information by refining that
$E_i$ into a sum $E_i = \sum_j E_{ij}$ where the $E_{ij}$ are rank one.

\begin{figure}[tbp]
\leavevmode
\epsfxsize=4in
\epsfbox{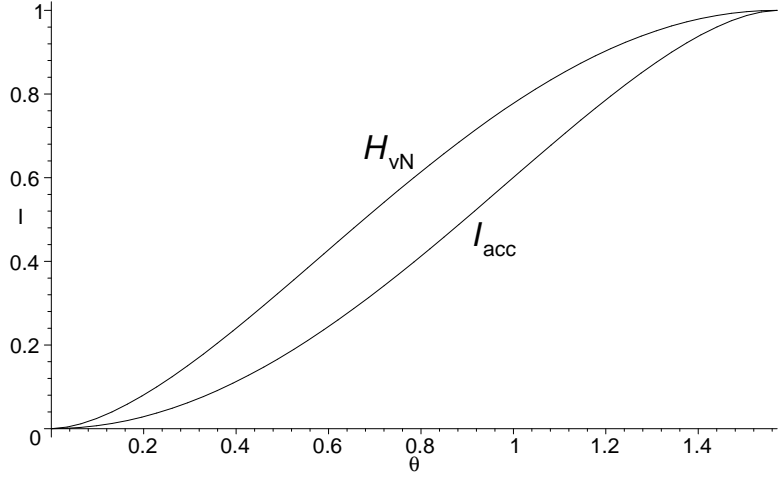}
\caption{A plot of the von Neumann entropy of the density matrix
and the accessible information for the ensemble of two pure quantum 
states with equal probabilities
and that differ by an angle of $\theta$, for $0 \leq \theta \leq \pi/2$.  
The top curve is the von Neumann entropy and the bottom the accessible 
information.
}
\label{graph-ai-vn}
\end{figure}

We now give some examples of the measurements maximizing accessible 
information.  The first is one of the simplest examples.  Suppose that we
have just two pure states in our ensemble, 
with probability $\frac{1}{2}$ each.  For  
example, we could take the states $\ket{\vpols}$ and $\ket{\ppols}$.  
Let us take $v_1 = (1,0)$ and $v_2 = (\cos \theta, \sin \theta)$.  
We will not prove it here, but
the optimal measurement for these is the von Neumann measurement
with two orthogonal vectors symmetric around $v_1$ and $v_2$.  That is,
the measurement with projectors
\begin{eqnarray}
w_1 &=& \big(\cos \textstyle(\frac{\pi}{2} + \frac{\theta}{2}), 
\sin (\frac{\pi}{2} + \frac{\theta}{2})\big)\\
w_2 &=& \big(\cos \textstyle(-\frac{\pi}{2} + \frac{\theta}{2}), 
\sin (-\frac{\pi}{2} + \frac{\theta}{2})\big)
\end{eqnarray}
This measurement is symmetric 
with respect to interchanging $v_1$ and $v_2$, and it leads to
a binary symmetric channel with error probability 
\begin{equation}
\cos^2 \left(\frac{\pi}{2} + 
\frac{\theta}{2}\right) = \frac{1}{2}-\frac{\sin \theta}{2}.
\end{equation}
The accessible information is thus $1-H_2(\frac{1}{2}-\frac{\sin\theta}{2})$.
Here $H_2$ is the Shannon entropy of a binary signal, i.e., 
\[
H_2(p) = - p \log p  - (1-p) \log (1-p)
\]
For the ensemble containing $v_1$ and $v_2$ with probability $\frac{1}{2}$
each, the density matrix is
\begin{equation}
\rho = \frac{1}{2}\left(\begin{array}{cc}
1+\cos^2\theta & \sin\theta\cos\theta\\
\sin\theta\cos\theta & 1 - \cos^2\theta
\end{array}
\right),
\end{equation}
which has eigenvalues $\frac{1}{2}\pm \cos \theta$, so the von Neumann
entropy of the density matrix is $H_2(\frac{1}{2}-\frac{\cos\theta}{2})$.
The values of $I_\mathrm{acc}$ and $H_\mathrm{vN}$ are plotted in 
Figure~\ref{graph-ai-vn}.  One can see that
the von Neumann entropy is larger than the accessible information.

Note that in our first example, the optimum measurement was a von Neumann 
measurement. If there are only two states in an ensemble, it has been 
conjectured that the measurement optimizing accessible information
is always a von Neumann measurement, in part because extensive
computer experiments have not found a counterexample \cite{Fuchs-Peres}.  
This conjecture 
has been proven for quantum states in two dimensions \cite{Levitinrecent}.  
Our next example shows that this conjecture does not hold for ensembles 
composed of three or more states.

Our second example is three photons with polarizations that differ by
$60^\circ$ each.  These are represented by the vectors 
\begin{eqnarray*}
v_0 &=& (1,0)\\
v_1 &=& \textstyle (-\frac{1}{2},\frac{\sqrt{3}}{2})\\
v_2 &=& \textstyle (-\frac{1}{2},-\frac{\sqrt{3}}{2})
\end{eqnarray*}
The optimal measurement for these states is the POVM corresponding to
the three vectors $w_i$ where $w_i \perp v_i$.  
We take $E_i = \frac{2}{3} w_iw_i^\dag$, in order for $\sum_i E_i = I$.
If we start with vector $v_i$, it is easy to see that we never
obtain $w_i$, but do obtain the other two possible outcomes with probability
$\frac{1}{2}$ each.  This gives an accessible information of
$I_\mathrm{acc} = \log 3 - 1$.  For
these three signal states, it is also easy to check that the density
matrix $\rho = \frac{1}{2}I$, so $H_\mathrm{vN}=1$.  Again,
we have $I_\mathrm{acc} < H_\mathrm{vN}$.

Given these two examples and some intuition, one might formulate the
conjecture
that $I_\mathrm{acc} \leq H_\mathrm{vN}$.
This is true, as in fact is a somewhat stronger theorem
which we will shortly state.
The first published proof of this theorem was given by 
Holevo \cite{holevobound}.
It was earlier conjectured by Gordon 
\cite{Gordon} and stated by Levitin with 
no proof \cite{Levitin69}.  

{\bf Theorem} (Holevo):
{\em
Suppose that we have a memoryless source emitting an ensemble of 
(possibly mixed) 
states $\sigma_i$, where $\sigma_i$ is emitted with probability $p_i$.  Let
\begin{equation}
\chi = H_\mathrm{vN}(\sum_i p_i \sigma_i) - \sum_i p_i H_\mathrm{vN}(\sigma_i).
\label{chi-definition}
\end{equation}
Then
\begin{equation}
I_\mathrm{acc} \leq \chi.
\label{Holevo-bound}
\end{equation}
}

The conditions for equality in this result are known.  
If all the $\sigma_i$ commute,
then they are simultaneously diagonalizable, and the situation is
essentially classical.  In this case, $I_\mathrm{acc}=\chi$;  
otherwise $I_\mathrm{acc}<\chi$.  

We define the $C_{1,1}$ capacity of a quantum channel as the maximum over
all ensembles of input states of the accessible information contained
by the corresponding ensemble of output states.  This is the capacity
of the channel for transmitting quantum information if we restrict 
the protocols that we use; namely, we only allow protocols that do not
send any states that are entangled over more than one channel use (this 
is the significance of the first `$1$' in the subscript), and
do not perform any joint quantum measurements involving more than one
channel output (this is the significance of the second `$1$'), and 
further we do not allow adaptive measurements of
the outputs (i.e., the measurement chosen cannot depend on results of
previous measurements on channel outputs).  Allowing adaptive measurements 
can in certain circumstances increase
the capacity, but they do not generally allow one reach the 
$C_{1,\infty}$ capacity discussed in the next section \cite{ping-pong}.

For example, if we consider the quantum channel where the sender
can choose to convey to the receiver either of the two pure quantum 
states of our first example, the optimum ensemble is the ensemble 
consisting of both states with equal probability, and the $C_{1,1}$
capacity is $1-H(\frac{1}{2}-\frac{\sin \theta}{2})$.  For the channel
which can convey to the receiver any of the three states in our 
second example, the maximum ensemble giving $C_{1,1}$ turns out to
be that which uses just two of the three states, each with probability
$\frac{1}{2}$.  This is our first example with $\theta=60^\circ$,
so this channel has a $C_{1,1}$ capacity 
of $1 - H(\frac{1}{2}-\frac{\sqrt{3}}{4}) \approx .6454$.

\section{The classical capacity of a quantum channel}

One can ask the question: is the $C_{1,1}$ capacity
the most information that can be sent per quantum state,
using only the three states of our second example?  The answer is, 
surprisingly, ``no''.  Suppose that we use the three length-two codewords 
$v_0\otimes v_0$, $v_1\otimes v_1$, and $v_2\otimes v_2$.  These are 
three pure states in the 
four-dimensional quantum space of two qubits.  Since there
are only three vectors, they lie in a three-dimensional subspace.  
The inner product between any two of these states is $\frac{1}{4}$.
One can show for this tensor product ensemble, the optimal accessible 
information is attained by using the von Neumann
measurement having three basis vectors obtained by ``pulling'' the
three vectors $v_i \otimes v_i$ apart until they are all orthogonal.  This
measurement gives $I_\mathrm{acc} = 1.369$ bits, larger
than twice the $C_{1,1}$ capacity, which is $1.2908$ bits.
We thus find that block coding and joint measurements
let us achieve a better information transmission rate than $C_{1,1}$.

Having found that length two codewords work better than length one codewords,
the natural question becomes: as the lengths of our codewords go to infinity,
how well can we do.  We define the $C_{1,\infty}$ capacity of a quantum
channel as the capacity over protocols which do not permit inputs 
entangled between two or more channel uses, but do allow joint quantum
measurements over arbitrarily many channel uses.  
A generalization of Shannon's 
giving the $C_{1,\infty}$ capacity has been proven.

{\bf Theorem} (Holevo\cite{Holevo98}, Schumacher--Westmoreland\cite{SW97}):
{\em The $C_{1,\infty}$ capacity of a quantum channel, i.e., that capacity 
obtainable using codewords composed of 
signal states~$\sigma_{i}$, where the probability of using $\sigma_i$ is $p_i$,
is 
\begin{equation}
\chi = H_\mathrm{vN}(\sum_i p_i \sigma_i) - \sum_i p_i H_\mathrm{vN}(\sigma_i).
\end{equation}
}
Note that $\chi$ is a function of the probabilistic ensemble of the
signal states $\{\sigma_i, p_i\}_i$ that we have chosen, where 
state $\sigma_i$ has $p_i$.  We will sometimes 
write $\chi(\{\sigma_i, p_i\}_i)$
so as to explicitly show this dependence.
Another approach to proving this theorem, which also provides some
additional results, appears in \cite{HaNa,OgNa}

We later give a sketch of the proof of the $C_{1,\infty}$
capacity formula in the special
case where the $\sigma_i$ are pure states.  We will first ask:
Does this formula give the true capacity of a quantum channel $\cal N$?
There are certainly protocols in which the sender, for example, uses
the two halves of an EPR pair of entangled qubits (as in Eq.~(\ref{EPR}))
as inputs for two separate
channel uses.  The question is: does allowing this type of protocol let 
one obtain a larger capacity?

Before we address this question (we will not be able to answer it)
we should give the mathematical description of a general quantum channel.
If ${\cal N}$ is a memoryless quantum communication channel, then
it must take density matrices to density matrices.  This means ${\cal N}$
must be a linear trace preserving positive map.  Here, linear is required
by the basic principles of quantum mechanics and trace preserving is 
required since the channel must preserve trace 1 matrices.  Positive
means the channel takes positive semidefinite matrices to 
positive semidefinite
matrices (and thus takes density matrices to density matrices).  
For ${\cal N}$ to be a valid quantum map, it must have one more 
property: namely, it must be completely positive.  This means that
${\cal N}$ is positive
even when it is tensored with the identity map.  There is a theorem 
\cite{Hellwig-Kraus}
that any linear completely positive map can be expressed as
\begin{equation}
{\cal N}(\rho) = \sum_i A_i \rho A_i^\dag,
\end{equation}
and the condition for the map to be trace-preserving is 
that the matrices $A_i$ 
satisfy $\sum_i A_i^\dag A_i = I$.

A natural guess at the capacity of a quantum channel ${\cal N}$ would
be the maximum of $\chi$ over all possible distributions of channel outputs,
that is,
\begin{equation}
\chi_\mathrm{max} ({\cal N}) = \max_{\{\sigma_i, p_i\}_i} 
\chi \big(\{{\cal N}(\sigma_i), p_i\}_i\big),
\end{equation}
since the sender can effectively communicate to the receiver any
of the states ${\cal N}(\sigma_i)$.  
We do not know whether this is
the capacity of a quantum channel; 
if the use of entanglement between separate inputs to the channel helps
to increase channel capacity,  it would be possible to exceed
this $\chi_\mathrm{max}$.
This can be addressed
by answering a question that is simple to state: Is $\chi_\mathrm{max}$
additive \cite{AHW, OsNa, King1, Shor-add, MSW, AB}?  
That is, if we have two quantum 
channels ${\cal N}_1$ and ${\cal N}_2$, is 
\begin{equation}
\chi_\mathrm{max} ({\cal N}_1 \otimes {\cal N}_2 ) = 
\chi_\mathrm{max} ({\cal N}_1)  +
\chi_\mathrm{max} ({\cal N}_2)  .
\label{additivityinchi}
\end{equation}
Proving superadditivity of the quantity $\chi_\mathrm{max}$ (i.e.,
the $\geq$ direction of Eq.~(\ref{additivityinchi}))
is easy.  The open question is
whether strictly more capacity can be attained by using the
tensor product of two channels jointly than by using them separately.

We now return to the discussion of the proof of the 
Holevo-Schumacher-Westmoreland theorem in the special case
where the $\sigma_i$ are pure states.  The proof of this case in fact
appeared before the general theorem was proved \cite{HJSWW96}.  
The proof uses three ingredients.  These are
(1) random codes, (2) typical subspaces, and (3) the square root measurement.

The square root measurement is also called the ``pretty good''
measurement, and we have already seen an example of it.
Recall our second example for accessible information,
 where we took the three vectors $v_i \otimes v_i$,
where $v_i = (\cos \frac{2\pi i}{3},\sin \frac{2\pi i}{3})$ for $i=0,1,2$.  
The optimal measurement for $I_\mathrm{acc}$ on
these vectors was the von Neumann measurement
obtained by ``pulling'' them farther apart until they were orthogonal.
This is an example of the square root measurement.

Suppose that we are trying to distinguish between vectors $w_1$, $w_2$,
$\ldots$, $w_n$, which appear with equal probability (the square root
measurement can also be defined for vectors having unequal probabilities,
but we do not need this case).  Let $\phi = \sum_i w_i w_i^\dag$.
The square root measurement has POVM elements
$E_i = \phi^{-1/2} w_i w_i^\dag \phi^{-1/2}$.  We have
\begin{equation}
\sum_i E_i = 
\phi^{-1/2} \left(\sum_i w_i w_i^\dag\right) \phi^{-1/2} = I,
\end{equation}
so these $E_i$ do indeed form a POVM.  

We can now give the coding algorithm for the capacity theorem for
pure states.
We choose $M$ codewords $u_j = v_{i_1} \otimes v_{i_2} \otimes \cdots 
\otimes v_{i_n}$,
where the $v_i$ are chosen at random with probability $p_i$.  We then
use these particular $M$ codewords $u_j$ to send information, where
the coding scheme is chosen so that each of these codewords is sent with
probability $\frac{1}{M}$.  The difficult part of the 
proof is now showing that a random codeword
can be identified with high probability.  

To decode, we perform the following steps:
\begin{enumerate}
\item Project into the typical subspace ${\cal T}$.
Most of the time, this projection works, and we obtain 
$w_j = (u^\dag \Pi_{\cal T} u_j)^{-1/2} \Pi_{\cal T} u_j$, 
where $\Pi_{\cal T}$ is the 
projection matrix onto the subspace ${\cal T}$.  
\item Use the square root measurement on the $w_j$.
\end{enumerate}
The probability of error, given that the original state was $w_j$, is
\begin{eqnarray*}
1-w_j^\dag E_j w_j &=& 1- w^\dag_j \phi^{-1/2} w_j w^\dag_j \phi^{-1/2} w_j \\
&=& 1- w_j^\dag \phi^{-1/2} w_j
\end{eqnarray*}
The overall probability of error is thus
\begin{equation}
1-\frac{1}{M} \sum_{j=1}^M |w_j^\dag \phi^{-1/2} w_j|^2.
\end{equation}
The intuition for why this procedure works (this intuition is apparently
not even close
to being rigorous, as the proof works along substantially different lines) 
is that for this probability of error to be small,
we need that $\phi^{-1/2}w_j$ is close to
$w_j$ for most $j$.  However, the $w_j$ are distributed 
more or less randomly in the typical subspace $\cal T$, so 
$\phi = \sum_j w_jw_j^\dag$ is moderately close to
the identity matrix on its support, and thus $\phi^{-1/2} w_j$ 
is close to $w_j$.  Note that we need that the number $M$ of
$u_j$ is less than $\dim {\cal T}$, or otherwise it would be impossible
to distinguish the $w_j$; as by Holevo's bound (\ref{Holevo-bound})
a $d$-dimensional
quantum state space can carry at most $d$ bits of information.   

\section{Calculating the $C_{1,\infty}$ capacity.}
\label{Holevo-calculation}
We now consider the problem of numerically finding the $C_{1,\infty}$
capacity.  Recall that this capacity was expressible as
\[
\max_{\{p_i, v_i\}_i} H_\mathrm{vN}({\cal N} ( \sum_i p_i v_i v_i^\dag)) 
- \sum_i p_i H_\mathrm{vN}({ \cal N} ( v_i v_i^\dag)) 
\]
where the maximum is taken over all ensembles $\{p_i, v_i\}_i$
of pure states in the input space of the channel.  We propose
to maximize this in stages, at first holding 
one parameter of the ensemble fixed, and then
holding other parameters of this ensemble fixed.

We first consider the problem of maximizing $C_{1,\infty}$ while
holding the average density matrix $\rho = \sum_i p_i v_i v_i^\dag$
fixed.  This is equivalent to the minimization problem
\begin{eqnarray}
{\mathrm{minimize\ \ }} && \sum_i p_i H_\mathrm{vN}({\cal N}(v_iv_i^\dag)) \nonumber \\
{\mathrm{subject\ to}} && \sum_i p_i v_i v_i^\dag = \rho.
\label{HolevoLP}
\end{eqnarray}
The problem of finding the probability distribution $p_i$ minimizing
the expression (\ref{HolevoLP} is a linear programming problem, 
albeit one with an
infinite number of variables $p_i$, one for each of the continuum of
pure states $v_i$.  There is a standard way of attacking such problems
called column generation.  
First, the linear program is solved with
$v_i$ restricted to be chosen from some fixed finite set of possible
signal states
(these correspond to columns of the linear program).  
We then find vectors $v$ which, if added to this linear program,
would yield a better solution.  By iterating the steps of finding
new vectors to add to the linear program and solving the resulting 
improved linear program, we
hope to eventually converge upon the right solution.  If we are
guaranteed to find a good vector $v_i$ to add if one exists, then
it turns out that we will indeed converge upon the optimal solution.

We now give a few more details of this process.  
If the vectors $v_i$ range over a $d$-dimensional input space,
there are $d^2$
constraints in this problem, arising from the degrees of freedom of
the matrix equality
\[
\sum_i p_i v_i v_i^\dag = \rho.
\]
(Both $v_iv_i^\dag$ and $\rho$ are $d \times d$ Hermitian matrices, 
yielding $d^2$ degrees of freedom.  
Note that $\sum_i p_i = 1$ is implicit in this matrix equality,
as this can be obtained by taking the trace of both sides.)
There is thus an optimum solution which has at most $d^2$ non-zero
values of $p_i$, one for each constraint of the problem.  

The success of column generation is dependent on how well we can find
a column which will be advantageous to add to the linear program.  To 
describe how to do this, we need to introduce the concept of the
dual of a linear program.

This dual is another linear program.  The constraints of the first 
program (called the primal program) correspond to the variables of the dual
program, and vice versa.  If the primal program is a minimization,
then the dual program is a maximization.  The fundamental theorem
of linear programming asserts that the optimum value of these two 
programs are equal.

We now give the dual of the program (\ref{HolevoLP}) above.  
The variables of this dual will be the entries of a Hermitian
matrix $\tau$.  The program is
\begin{eqnarray}
\nonumber
{\mathrm{maximize\ \ }} && \Tr \, \tau \rho \\
{\mathrm{subject\ to}} && v^\dag \tau v \leq H_\mathrm{vN}({\cal N}(v v^\dag)) 
{\mathrm{\ \ \ for\ all\ unit\ vectors\ }}v \in {\mathbb{C}}^d.
\label{dualHolevoLP}
\end{eqnarray}
It may be instructive to consider 
the proof that the optimal value of the primal is greater
than or equal to the optimal value of the dual.  Suppose we have a 
$\tau$ such that $\Tr\, \tau \rho = {\bf x}$. Then
\begin{eqnarray}
\nonumber
\sum_i p_i H_\mathrm{vN}({\cal N}(v_iv_i^\dag)) &\geq& 
\sum_i p_i v_i^\dag \tau v_i \\
\nonumber
&=& \sum_i p_i \Tr\,\tau v_i v_i^\dag \\
&=& \Tr\,\tau \rho \ =\  {\bf x}.
\label{dual-primal-ineq}
\end{eqnarray}
Suppose we have an optimal solution of the primal program restricted
to a given fixed set of variables $v_i$.  This solution will correspond
to a dual problem with some optimal Hermitian matrix $\tau$.  
To find a good column to add to the primal, we need to find a $v$ 
that violates the constraints of the dual, i.e., such that 
\begin{equation}
\label{nonlinopt}
  H_\mathrm{vN}({\cal N}( v v^\dag)) - v^\dag \tau v < 0
\end{equation}
This is a nonlinear optimization problem which I
believe should be solvable in low dimensions
by gradient descent.  One can look
for good vectors to add by starting at an arbitrary vector $v$ and
proceeding downhill to find local minima of the expression~(\ref{nonlinopt}). 
One should start both at random points, and at the  
vectors $v_i$ having nonzero probability in the current solution (these
latter will improve the solution by adapting to the perturbation made
to the problem since the last iteration).
For high dimensions,
this technique will grow inefficient exponentially fast, but
the curse of dimensionality also may mean that the
problem is intrinsically difficult.  

We now need to show how to change the average density matrix $\rho$
of our ensemble so as to improve the optimal value.  Recall that
we want to
maximize 
\begin{equation}
H_\mathrm{vN}({\cal N}( \rho)) - \sum_i p_i H_\mathrm{vN}({\cal N}(v_i v_i^\dag)).
\label{needtomax}
\end{equation}
By the linear programming duality~(\ref{dual-primal-ineq}), 
this is smaller than
\begin{equation}
H_\mathrm{vN}({\cal N}( \rho)) - \Tr\,\rho \tau,
\label{maximizerho}
\end{equation}
for an arbitrary ensemble $\{p_i, v_i\}_i$, 
and equal at the current maximum.  From the concavity
of entropy, the expression (\ref{maximizerho}) is concave in $\rho$.
Thus, if there is no direction to change $\rho$ that will increase 
the maximum (\ref{maximizerho}), there is also no direction that will 
increase (\ref{needtomax}), and we have found the optimal value of $\rho$ 
(at least for our current set of $v_i$).  If there is a direction that 
increases (\ref{maximizerho}), then we can use binary search to find
the optimum distance to move $\rho$ in that direction.  
For the complete linear program with a continuum of variables ${v_i}$, 
we can use a
smoothness argument to show that this same direction will also increase
the objective function.  This argument, unfortunately, does not
appear to carry over to the finite dimensional linear program on a
fixed set of $v_i$ that we actually are solving.  If things work well,
it may turn out that attempting to move in this direction will result
in a procedure that always converges to the optimum.  Otherwise, we may
have to use the polyhedral structure of the solution to our finite
dimensional linear program to discover a good direction to move $\rho$.

The derivatives for Eqs.~(\ref{nonlinopt}) and (\ref{maximizerho})
can both be calculated explicitly.  This can be done using the 
estimate
\[
H(\rho + \epsilon \Delta) = H(\rho - \epsilon \Tr \Delta \log \rho 
+ O(\epsilon^2)
\]
which holds for matrices with $\Tr \rho = 1$ and $\Tr \Delta = 0$.
It can be derived from the integral expression for 
$\log(\rho + \epsilon \delta)$ given in Eq.~(20) of \cite{Ruskai}.

We thus propose to find the value of $C_{1,\infty}$ numerically by
iterating the following steps
\begin{enumerate}
\item With a fixed set of $v_i$, and the constraint 
$\sum_i p_i v_i v_i^\dag = \rho$, solve the linear program~(\ref{HolevoLP}).
\item Change $\rho$ so as to improve the value of the linear program
solution above.
\item Find a quantum states $v$ corresponding to columns it would
be advantageous to add to the linear program.
\end{enumerate}
If none of the steps can improve the solution, then we have 
discovered the optimum value of $C_{1,\infty}$.   The hard step is (3);
this is a non-linear optimization problem for which we have no good
criteria to test whether we have discovered the global optimum.   In
large dimensions, this will clearly be the bottleneck step in making the 
procedure impractical.

\section{Calculating the $C_{1,1}$ capacity.}
We now very briefly describe how the ideas of the last section
might be used to give a heuristic procedure for finding the $C_{1,1}$
capacity.  Our technique again uses column generation.
This time we propose to alternate between optimizing the
ensemble of signal states used for the input, and optimizing the
measurement used by the receiver.  This is not guaranteed to find the
global optimum; the second example of section~\ref{section-examples}
has four different local optima that are stable points in this procedure;
the three ensembles each containing two of the signal states with 
probability $\frac{1}{2}$, and the ensemble containing all three with 
probability $\frac{1}{3}$ \cite{ping-pong}.  

To find the optimal signal states, given a fixed measurement, we can
in fact use exactly the procedure given in the previous section.  By
fixing the measurements, we have defined a quantum channel, with the input
being a quantum state and the output being the results of the measurement,
and so the procedure of the previous section is applicable.
This is not an arbitrary quantum channel, as the output state
is classical.  Unfortunately, at this point we do not see how to use this
fact to simplify the calculation of the capacity.

Finding the optimal measurement, given the signal states, can be
done using essentially the same ideas as in the previous section.  
The measurement we use must extract a maximal amount of information,
and so we can assume that each $E_i$ takes the form $E_i = q_i w_i w_i^\dag$
for some unit quantum state vector $w_i$ in the output space.   The
condition that these form a POVM is
\[
\sum_i q_i w_i w_i^\dag = I.
\]

Now, the expression for the capacity is the entropy of the input
minus the entropy of the input, given the output.  This can be seen
to be linear in $q_i$.  Again, we have an infinite dimensional linear 
program, which we can solve using the technique of column generation.
In this case, it is even slightly simpler; because the constraints 
are $\sum_{i} q_i w_i w_i^\dag = I$, we do not need to incorporate any 
additional steps of optimizing $\rho$.  

\section{Entanglement-assisted capacities}

In this section, we define the entanglement-assisted capacity of a 
quantum channel, and give the expression for it.  For motivation, 
we first describe two surprising phenomena in quantum communication:
{\em superdense coding} and {\em quantum teleportation.} 

\begin{figure}[tbp]
\leavevmode
\epsfxsize=4in
\epsfbox{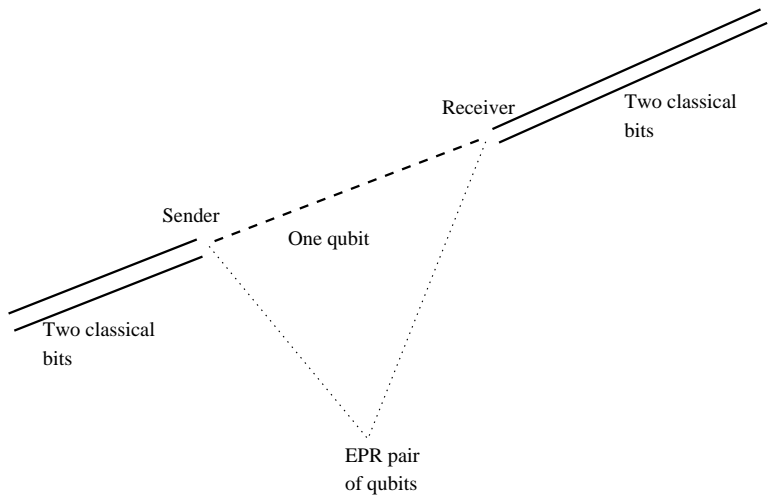}
\caption{A schematic drawing of superdense coding.  The sender
can communicate two classical bits to the receiver using one qubit
and a shared EPR pair.  Here, the sender makes the same unitary transformation
that the receiver would make in quantum teleportation, and the receiver 
makes the joint measurement that the sender would make in quantum 
teleportation.}
\label{superdense}
\end{figure}

The process of {\em superdense coding} 
uses a shared EPR pair and a single qubit to encode two classical 
bits \cite{sdcoding}.  This is an improvement on the capacity
of a noiseless, unassisted, quantum channel, which takes one 
quantum bit to send a classical bit.
We will assume that the shared EPR pair is in the state
\[
\frac{1}{\sqrt{2}} \left( \ket{01} - \ket{10} \right)
\]
where the sender holds the first qubit and the receiver holds the second
qubit.
In this protocol, the sender starts by taking an EPR pair and applying
to it either the identity operation or one of the three Pauli matrices
\[
\sigma_x = \left( \begin{array}{cc}
0&1 \\
1&0
\end{array} \right),\qquad
\sigma_y = \left( \begin{array}{cc}
0&-i \\
i&\phantom{-}0
\end{array}\right),\qquad
\sigma_z = \left( \begin{array}{cc}
1& \phantom{-}0\\
0&-1
\end{array}\right).
\]
He then sends his qubit to the receiver.  The receiver now holds one
of the four quantum states
\begin{eqnarray*}
\frac{1}{\sqrt{2}} \left( \ket{01} - \ket{10}  \right),&&
\frac{1}{\sqrt{2}} \left( \ket{11} - \ket{00}  \right),\\
\frac{i}{\sqrt{2}} \left( \ket{11} + \ket{00}  \right),&&
\frac{1}{\sqrt{2}} \left( \ket{01} + \ket{10}  \right). 
\end{eqnarray*}
These four states are known as the Bell basis, and they are mutually
orthogonal.  The receiver can thus uniquely identify which of these 
states he has, and so can unambiguously identify one of four messages,
or two bits.
(See Figure \ref{superdense}.)

\begin{figure}[tbp]
\leavevmode
\epsfxsize=4in
\epsfbox{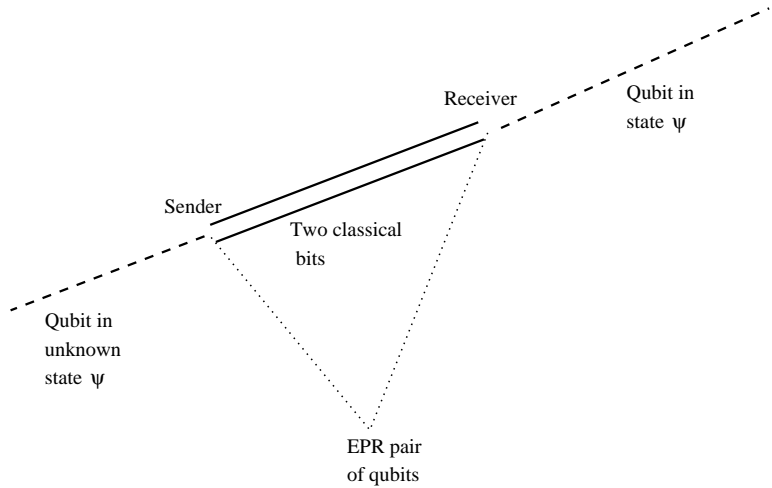}
\caption{A schematic drawing of quantum teleportation.  The sender
has a qubit in an unknown state $\psi$ that he wishes to send to the 
receiver.  He also has half of an EPR state which he shares
with the receiver.
The sender makes a joint measurement on the unknown qubit and half of
his EPR state, and communicates the results (2 classical bits) 
to the receiver.  The receiver then makes one of four unitary transformations
(depending on the two classical bits he received)
on his half of the EPR state to obtain the state $\psi$.}
\label{teleport}
\end{figure}

There is a converse process to superdense coding known as quantum
teleportation. 
It is impossible to send a quantum
state over an unassisted classical channel.  However,
quantum teleportation lets a sender and a receiver who share an
EPR pair of qubits communicate one qubit by sending two classical bits 
and using this EPR pair
\cite{teleportation}.  
(See Figure \ref{teleport}.) 
In quantum teleportation, the sender measures the unknown quantum state
state and the EPR pair in the Bell basis, and sends the receiver
the two classical bits which are the results of this measurement.
The receiver then performs a unitary operation.  The measurements
the sender makes are the same ones the receiver makes in superdense coding,
and the unitary transformations the receiver performs are those the
sender performs in superdense coding.

Quantum teleportation is a counterintuitive process, which at first sight 
seems to violate certain laws of physics; however, 
upon closer inspection one discovers that no actual paradoxes 
arise from teleportation.  Teleportation cannot be used for
superluminal communication, because the classical bits must travel at
or slower than
the speed of light.  A continuous quantum state, which at first sight
appears to contain many more than two bits of information, appears to have 
been transported using two discrete bits; however, by Holevo's bound, 
Eq.~(\ref{Holevo-bound}), one qubit can be used to transport at most one 
classical bit of information, so it is not possible to increase the capacity 
of a classical channel by encoding information in a teleported qubit.
Finally, there is a theorem of quantum mechanics that an unknown quantum
state cannot be duplicated \cite{nocloning}.  However, this no-cloning
theorem (as it is known) is not violated; the original
state is necessarily destroyed by the measurement, so teleportation cannot
be used to clone a quantum state.  

We now give another capacity for quantum channels, one which has a 
capacity formula which can actually be completely proven, even in the
case of infinite-dimensional Hilbert spaces 
\cite{CEshort,CElong,HolEA1,HolEA2}.  
Recall that if ${\cal N}$ is a noiseless
quantum channel, and if the sender and receiver possess shared
EPR pairs, they can use superdense coding to double the classical
information capacity of ${\cal N}$.  Similarly, if ${\cal N}$ is a noiseless
classical channel, EPR pairs can increase the capacity of the channel to 
send quantum information from zero qubits to half a qubit per channel use.

In general, if ${\cal N}$ is a noisy quantum 
channel, using shared EPR pairs can increase both the classical and
quantum capacities of
${\cal N}$.  With the aid of entanglement, the capacity for sending quantum
information becomes exactly half of the capacity for classical
information (this is a direct consequence of the phenomena of superdense
coding and teleportation).  
We define the entanglement assisted capacity, $C_E$, 
as the quantity
of classical information that can asymptotically be sent per channel use
if the sender and receiver have access to a sufficient quantity of shared
entanglement.

{\bf Theorem} (Bennett, Shor, Smolin, Thapliyal \cite{CEshort,CElong}):
{\em
The entanglement assisted capacity is
\begin{equation}
C_E({\cal N}) = \max_{\rho \in {\cal H}_{\mathrm{in}}} 
H_{\mathrm{vN}} (\rho)
+H_{\mathrm{vN}} ( ({\cal N} (\rho))
-H_{\mathrm{vN}} ( ({\cal N} \otimes {\cal I}) (\Phi_\rho))
\label{eac}
\end{equation}
where $\rho \in {\cal H}_\mathrm{in}$ is a density matrix over the
input space.
Here, $\Phi_\rho$ is a pure state over the tensor product space
${\cal H}_\mathrm{in} \otimes {\cal H}_R$
such that $\Tr_R \Phi_\rho = \rho$.
Here ${\cal H}_\mathrm{in}$
is the input state space and ${\cal H}_R$ is a reference system.
The third term of the right hand side of~(\ref{eac}),
$H_{\mathrm{vN}} ( ({\cal N} \otimes {\cal I}) (\Phi_\rho))$, 
is the entropy of the state resulting after the first half of $\Phi_\rho$ 
is sent through the channel ${\cal N}$, while the identity operation
is applied to second half.
The value of this term
is independent
of which reference system ${\cal H}_R$ and which pure state $\Phi_\rho$
are chosen.
}

The quantity being minimized in the above formula (\ref{eac}) is 
sometimes called 
quantum mutual information, and it is a generalization of the
expression for mutual information
in the form of Eq.~(\ref{input-plus-output-less-joint}).
The proof of this result uses
typical subspaces,
superdense coding,
the Holevo-Schumacher-Westmoreland theorem on the classical capacity 
of a quantum channel, and
the strong subadditivity property of von Neumann entropy.
The entanglement assisted capacity is a convex function of $\Tr_S \rho$, and 
so can be maximized with a straightforward application of gradient
descent.

In the entanglement-assisted capacity, the communication protocol 
consumes the resource of entanglement.  In general, it takes 
$H_\mathrm{vN} (\rho)$ bits of entanglement (i.e., EPR pairs) per channel
use to achieve the capacity $C_E({\cal N})$ in Eq.~(\ref{eac}).  I have
also conjectured a formula for the capacity of a quantum channel 
using protocols which are only allowed to use a limited amount of 
entanglement.  This formula is as follows.

{\bf Conjecture:} 
{\em
If the available entanglement per channel use is $B$ bits,
the capacity available is 
\begin{equation}
 \max_{\rho_i, p_i \atop
\sum_i p_i H(\rho_i) \leq B} 
\sum_i p_i H_{\mathrm{vN}} (\rho_i)
+H_{\mathrm{vN}} ( ({\cal N} (\sum_i p_i \rho_i))
-\sum_i p_i H_{\mathrm{vN}} ( ({\cal N} \otimes {\cal I}) (\Phi_{\rho_i})).
\label{limited-eac}
\end{equation}
where $\Tr_2 \Phi_{\rho_i} = \rho_i$, as in Eq.~(\ref{eac}). 
Here, the maximization is over all probabilistic ensembles of density matrices
$\{\rho_i, p_i\}_i$ where $\rho_i \in {\cal H}_\mathrm{in}$, 
$\sum_i p_i = 1$, and the average entropy of the ensemble,
$\sum_i p_i H_\mathrm{vN}(\rho_i)$, is at most $B$.}  

\noindent
I have a 
protocol which achieves this bound, and I can prove a matching upper bound 
over a restricted class of protocols.  

A heuristic for finding this capacity with assistance by limited entanglement 
can be constructed using linear programming, column generation, and 
non-linear optimization, 
along the same lines as the protocol of Section \ref{Holevo-calculation}.
An extra constraint must be added that bounds the average entropy of the 
ensemble.
One additional difficulty will be that the 
non-linear optimization problem needed to find good columns to add
appears to become substantially harder.

\section{Sending Quantum Information}

Finally, we briefly mention the problem of sending quantum information
(i.e., a quantum state) over a noisy quantum channel.  In this scenario,
several of the theorems that make classical channel capacity behave
so nicely are demonstrably not true.  Here, a feedback channel from the
receiver to the sender, or a classical two-way side channel,
will increase the quantum
channel capacity, leading to several different
capacities for transmitting quantum information. 
For the two-way quantum capacity,
$Q_2$, the sender and receiver have a classical side channel they
can use for free,
For the quantum capacity with feedback,
$Q_{FB} < Q_2$, the receiver has a classical feedback channel 
from himself to 
the sender.
For the one-way quantum capacity, $Q\leq Q_{FB}$, 
all communication is directly from 
the sender
to the receiver over the noisy quantum channel ${\cal N}$.  
The quantities $Q_2$ is closely related to a quantity defined on quantum
states called the distillable entanglement \cite{Bruss}.  Despite
substantial study, not only do we not have any good ways to compute either
$Q_2$ or $Q_{FB}$,
but we also have no simple capacity formulas for representing it.
There is a 
capacity formula for the one-way quantum capacity $Q$.  
It is essentially the last two terms of the expression (\ref{eac})
for entanglement-assisted capacity 
\begin{equation}
Q({\cal N}) = \lim_{n \rightarrow \infty} \frac{1}{n}
\max_{\rho \in {\cal H}_{\mathrm{in}}} 
H_{\mathrm{vN}} ({\cal N}^{\otimes n}(\rho))
-H_{\mathrm{vN}} ( ({\cal N}^{\otimes n} \otimes {\cal I}) \Phi_\rho)
\label{coherent-info}
\end{equation}
where $\rho$, ${\cal H}_{\mathrm{in}}$ and ${\Phi_\rho}$ are defined as 
in (\ref{eac}).  The
quantity being maximized (before the limit $n \rightarrow \infty$) 
is called the {\em coherent information}.
We need to take the maximum over the tensor product of $n$ uses of the
channel, and let $n$ go to infinity, because unlike the classical 
(or the quantum)
mutual information, the coherent information is not additive \cite{dVSS}.
The quantity Q({\cal N}) (\ref{coherent-info}) is the quantum
capacity of a noisy 
quantum channel ${\cal N}$ \cite{ci-lloyd,ci-BST,ci-hl,ci-shor}. 
Even for maximizing the single-symbol expression (that is, taking $n=1$ in
Eq.~(\ref{coherent-info})), the calculation of the coherent information 
appears to be a difficult optimization problem, as there may be multiple
local maxima.  It would be a significant accomplishment to discover a
good means of calculating this; unfortunately, I do not have any useful
suggestions.

\section*{Acknowledgement}
I would like to thank Mary Beth Ruskai for encouraging me to work
on the question of calculating quantum channel capacities, as well
as for her help with some the technical issues in quantum channel
capacity.

\setlength{\itemsep}{0pt}
\setlength{\parskip}{0pt}
\setlength{\parsep}{0pt}

\end{document}